\documentclass{aa}  

\usepackage{graphicx}
\usepackage[varg]{txfonts}
\usepackage{natbib,twoopt}
\usepackage{amsmath}
\usepackage[nopatch]{microtype}
\usepackage{booktabs}
\usepackage{multirow}
\usepackage{hyphenat}
\usepackage{ulem}
\usepackage{amssymb}
\usepackage{threeparttable}
\usepackage{color,soul}
\usepackage{lscape,rotating}
\usepackage{xcolor}
\usepackage[colorlinks=True,allcolors=blue]{hyperref}
\usepackage{breakurl}

\def\asec{\ifmmode ^{\prime\prime}\else$^{\prime\prime}$\fi}
\def\msun{M$_{\odot}$}
\def\rsun{R$_{\odot}$}

\def\degs{\ifmmode ^{\circ}\else$^{\circ}$\fi}
\def\amin{\ifmmode ^{\prime}\else$^{\prime}$\fi}
\def\asec{\ifmmode ^{\prime\prime}\else$^{\prime\prime}$\fi}
\def\fss{\hbox{$.\!\!^{\rm s}$}}        % Fractions of seconds
\def\fdg{\hbox{$.\!\!^\circ$}}          % Fractions of degrees
\def\farcs{\hbox{$.\!\!^{\prime\prime}$}}  % Fractions of arcseconds
\def\h{$^{\rm h}$}
\def\m{$^{\rm m}$}

\def\psr{J1544}
\def\erossrc{4eRASS J154415.4$-$255531}
\def\fgl{4FGL J1544.2$-$2554}
\def\ergs{erg~s$^{-1}$}

\def\ps{Pan-STARRS}
\def\fermi{\textit{Fermi}}
\def\gaia{\textit{Gaia}}

\newcommand{\flux}{erg~s$^{-1}$~cm$^{-2}$}
\newcommand{\kirr}{erg~cm$^{-2}$~s$^{-1}$~sr$^{-1}$}

\begin{document}

\titlerunning{\fgl: a new spider pulsar candidate} 
\authorrunning{Karpova et al.}

   \title{\fgl: a new spider pulsar candidate }

   \author{A. V. Karpova\thanks{E-mail: annakarpova1989@gmail.com}\inst{1},
          S. V. Zharikov\inst{2},
          D. A. Zyuzin\inst{1},
          A. Yu. Kirichenko\inst{2,1},
          Yu. A. Shibanov\inst{1}
          \and
          I. F. M\'arquez\inst{2}
          }

   \institute{Ioffe Institute, 26 Politekhnicheskaya, St. Petersburg, 194021,  Russia 
    \and
        Instituto de Astronom\'ia, Universidad Nacional Aut\'onoma de M\'exico, Apdo. Postal 877, Baja California, M\'exico, 22800
             }

   \date{Received ..., 2024; accepted ...}

% \abstract{}{}{}{}{} 
% 5 {} token are mandatory
 
  \abstract
   {Spider pulsars are millisecond pulsars in tight binary systems, in which  a low-mass companion star is heated and ablated by the pulsar wind. 
   Their observations allow one to  study stellar evolution with formation of millisecond pulsars and physics of superdense matter in neutron stars. 
   However, spiders  are  rare due to difficulties of their discovery  using typical radio search techniques.
   The \fermi\ $\gamma$-ray source \fgl\ was recently proposed as a pulsar candidate, and its likely X-ray and optical counterparts  with  the galactic coordinates $l\approx344\fdg76$, $b\approx22\fdg59$ 
   and the magnitude $G\approx20.6$ were found  using the \textit{eROSITA} and \gaia\ surveys. 
}  
   {Our goals are to  study whether the source is a new spider pulsar and to estimate its fundamental parameters. 
   }
   {We  performed the first optical time-series multi-band photometry of the object. 
   We used the Lomb-Scargle periodogram to search for its brightness  periodicity and fitted its light curves with a model of direct heating of the binary companion by the pulsar wind.}
   {The source shows a strong brightness variability with  a  period of $\approx$2.724 h and an amplitude of $\gtrsim$2.5 mag, and its light curves  have a single broad  peak per period.
   These  features are  typical for spider pulsars.  
   The  curves are well fitted by the direct heating model, resulting in  an orbit inclination of  the presumed spider system of $\approx 83$\degs, a companion mass of $\approx 0.1$~\msun, its ``day-side'' and ``night-side'' temperatures of $\approx 7200$~K and $\approx 3000$~K, a Roche-lobe filling factor of $\approx 0.65$ and a distance of $\approx 2.1$~kpc. 
   }
   {Our findings suggest that \fgl\ is a spider pulsar.    
   This  encourages searches for the pulsar millisecond pulsations in the radio and $\gamma$-rays to confirm its nature.}


   \maketitle
%
%-------------------------------------------------------------------
\section{Introduction}
\label{sec:introduction}

%%%%%%%%%%%%%%%%%%%%%%% Fig. PSR fieled %%%%%%%%%%%%%%%%%%%%%%%%%%%%%%%%
\begin{figure*}
\begin{minipage}[h]{1.\linewidth}
\center{
\includegraphics[width=0.4\linewidth,clip=]{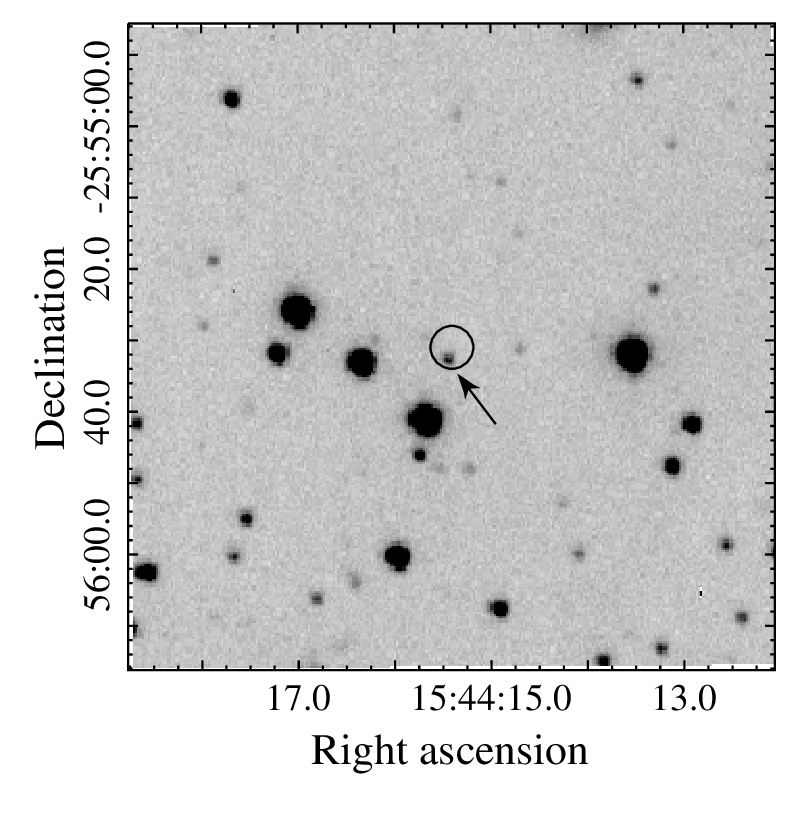}
\includegraphics[width=0.4\linewidth, clip=]{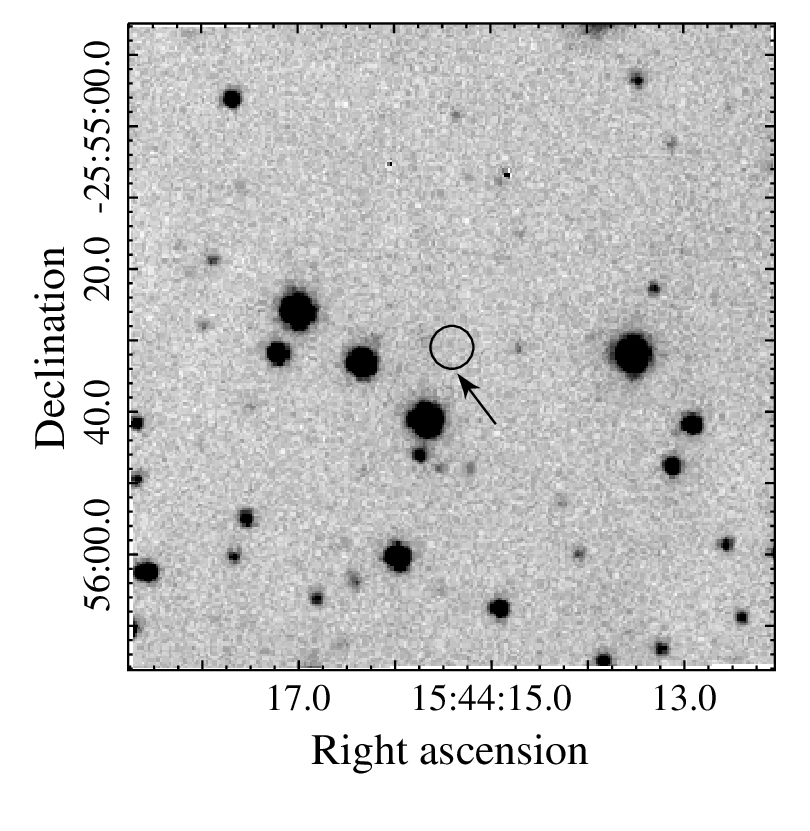}
}
\end{minipage}
\caption{1\farcm5 $\times$ 1\farcm5 individual $R_c$-band images of the \psr\ field obtained with the 2.1-m OAN-SPM telescope near the maximum (left) and the minimum (right) brightness phases. The optical counterpart candidate is shown with the arrow. 
The circle with a radius of 3\asec\ indicates the 1$\sigma$ position uncertainty of \erossrc\ proposed as the X-ray counterpart \citep{mayer2024}. 
}  
\label{fig:img}
\end{figure*}
%%%%%%%%%%%%%%%%%%%%%%% Fig. PSR fieled %%%%%%%%%%%%%%%%%%%%%%%%%%%%%%%%

At the moment, 146 millisecond pulsars (MSPs) have been detected by the \fermi\ Gamma-ray Space Telescope since its launch in 2008 \citep{fermipulsars}. 
A considerable number of them are so-called ``spider'' pulsars, which are divided into the ``redback'' (RB) and ``black widow'' (BW) subclasses \citep{roberts2013,manchester2017}.
These are close binary systems with orbital periods of $\lesssim$1~d in which the face side of a tidally locked low-mass companion star is heated and ablated by the high-energy pulsar wind. 
BWs have degenerate companions with masses of $\lesssim$0.05~\msun, while companions of RBs are non-degenerate and heavier (0.1--1~\msun). 
In such systems, masses of neutron stars (NSs) can exceed 2~\msun, which is considerably higher than the canonical NS mass of 1.4~\msun\ \citep[e.g.][]{nieder2020,romain2022,Thongmeearkom2024}. 
This makes  the studies of spiders particularly interesting for constraining  the still poorly known properties of super-dense matter inside NSs.

Until now, about 20 RBs and 40 BWs have been discovered in the Galactic disc \citep{strader2019,swihart2022} and a similar amount -- in globular clusters \citep{psrs-in-gcs}.  
Unfortunately, the material evaporated from a companion often obscures the pulsar radio emission for the part of the orbit, which makes it difficult to detect pulsations and, therefore, to find new spider systems in radio surveys. 
Nevertheless, identification is still possible through multi-wavelength investigations of likely counterparts to the \fermi\ unassociated sources, especially in the optical and X-rays \citep[e.g.][]{salvetti2017}. 
Such searches have resulted in the discoveries of several promising spider pulsar candidates \citep[e.g.][]{li2021,swihart2021,swihart2022,halpern2022,karpova2023,zyuzin2024}.

The unassociated \fermi\ source \fgl\ (hereafter \psr) was recently proposed as  a MSP candidate based on its $\gamma$-ray spectral and timing properties, with a rather high probability of $0.75$ \citep{mayer2024}.
Using the data obtained  by the extended ROentgen Survey with an Imaging Telescope Array (\textit{eROSITA}) aboard the Spektrum-Roentgen-Gamma (SRG) space observatory, the authors found a likely X-ray counterpart to \psr, \erossrc. 
The counterpart spatially coincides  with a star (\gaia\ catalogue ID 6235002859670996352), which could be a binary companion of the proposed MSP candidate (see  figure~7 in \citealt{mayer2024}).
According to the \gaia\ Data Release 3 (DR3) catalogue, its magnitude is $G=20.6$ and the coordinates are R.A.(2016) = 15\h44\m15\fss455 and Dec.(2016) = $-$25\degs55\amin32\farcs688 \citep{gaia2016,gaia-dr3}.

We also found this companion candidate in the Panoramic Survey Telescope and Rapid Response System survey \citep[\ps;][] {flewelling2020}, Zwicky Transient Facility \citep[ZTF;][]{ztf} and Legacy Surveys DR 10 \citep{legacy} catalogues.
In these surveys the candidate shows a strong variability at a level of  $\ga 1$~mag. 
Combined with the detected X-ray and $\gamma$-ray emission, such a strong optical variability implies that  \psr\ is a spider pulsar. 
In this case, its optical brightness  is expected to  be modulated with the orbital period of the binary system.    
However, the available optical data are too sparse and noisy to be used for a periodicity search. 

To clarify the nature of this source, we thus carried out the dedicated time-series optical photometric observations. 
Here we report the period of the system and estimate its parameters based on the light curve analysis. 
We conclude that \psr\ is indeed a likely member of the spider class. 
 
The paper is organised as follows: observations and data reduction are described in Sec.~\ref{sec:data}, data analysis -- in Sec.~\ref{sec:lc-mod}, discussion -- in Sec.~\ref{sec:discussion} and conclusions -- in Sec.~\ref{sec:conclusion}.

%%%%%%%%%%%%%%%%%%%%%%%%%%%%%%%%%%%%%%%%%%%%%%%%%%%%%%%%%%%%%%%
%%%%%%%%%%%%%%%%%%%%%%%%%%%%%%%%%%%%%%%%%%%%%%%%%%%%%%%%%%%%%%%
%%%%%%%%%%%%%%%%%%%%%%%%%%%%%%%%%%%%%%%%%%%%%%%%%%%%%%%%%%%%%%%
%%%%%%%%%%%%%%%%%%%%%%%%%%%%%%%%%%%%%%%%%%%%%%%%%%%%%%%%%%%%%%%

\section{Observations and data reduction}
\label{sec:data}

We performed time-series photometric observations of the \psr\ field in the Johnson-Cousins $B$, $V$, and $R_c$ bands with the ``Rueda Italiana'' instrument attached to the 2.1-m telescope at the Observatorio Astron\'omico Nacional San Pedro M\'artir (OAN-SPM) in Mexico on  June 8--13, 2024. 
The weather conditions were either photometric or clear with an effective seeing between 1\farcs5 to 2\farcs1 (derived as the full width at the half maximum of a point source spatial profile in an image).
All images were taken with  exposures of 400~s. 
The log of observations is given in Table~\ref{log}.  
Standard data reduction of raw frames was carried out using the {\sc IRAF} package.
To calculate the astrometric solution, we used  a set of  reference stars from the \gaia\ DR3 catalogue. 
The formal $rms$ uncertainties of the resulting astrometric fit were $\Delta$R.A.~$\lesssim$~0\farcs023 and $\Delta$Dec.~$\lesssim$~0\farcs027. 
The Landolt standards  PG 1323 and PG 1657  \citep{1992AJ....104..340L} were observed for the photometric calibration. 
We employed the aperture photometry to extract the target light curves, and a differential technique was applied to eliminate the variations due to the changing weather conditions,  using three non-variable bright field stars as references.  

%%%%%%%%%%%%%%%%%%%%%%%%%%%%%%%%%%%%%%%%%%%%%%%%%%%%%%%%%%%%%%%
%%%%%%%%%%%%%%%%%%%%%%%%%%%%%%%%%%%%%%%%%%%%%%%%%%%%%%%%%%%%%%%
%%%%%%%%%%%%%%%%%%%%%%%%%%%%%%%%%%%%%%%%%%%%%%%%%%%%%%%%%%%%%%%
%%%%%%%%%%%%%%%%%%%%%%%%%%%%%%%%%%%%%%%%%%%%%%%%%%%%%%%%%%%%%%%

%%%%%%%%%%%%%%%%%%%%%% Table Observations %%%%%%%%%%%%%%%%%%%%%
\begin{table}
\centering{
\caption{Log of the \psr\ observations.} 
\label{log}
\begin{tabular}{cllc}
\hline
Date       & Filters   & Number  of   &  Duration \\
2024, June &           & exposures    &  (hours) \\ 
\hline
08         & $R_c$     & 21           & 2.3 \\    
09         & $V, R_c$  & 10, 10       & 3.1 \\
10         & $R_c$     & 3            & 0.3 \\
11         & $B,~R_c$  & 6, 17        & 4.0 \\
12         & $B,~R_c$  & 6, 6         & 1.3 \\ 
13         & $V,~R_c$  & 18, 17       & 4.0 \\ 
\hline
\end{tabular}}
\end{table}
%%%%%%%%%%%%%%%%%%%%%% Table Observations %%%%%%%%%%%%%%%%%%%%%

\section{Data analysis and results}
\label{sec:lc-mod}

Examples of the images of the \psr\ field obtained in the $R_c$ band are presented in Fig.~\ref{fig:img}. 
They demonstrate that the likely optical counterpart to the $\gamma$-ray source is strongly variable.

\begin{figure}
\begin{minipage}[h]{1.\linewidth}
\includegraphics[width=1.\linewidth, clip]{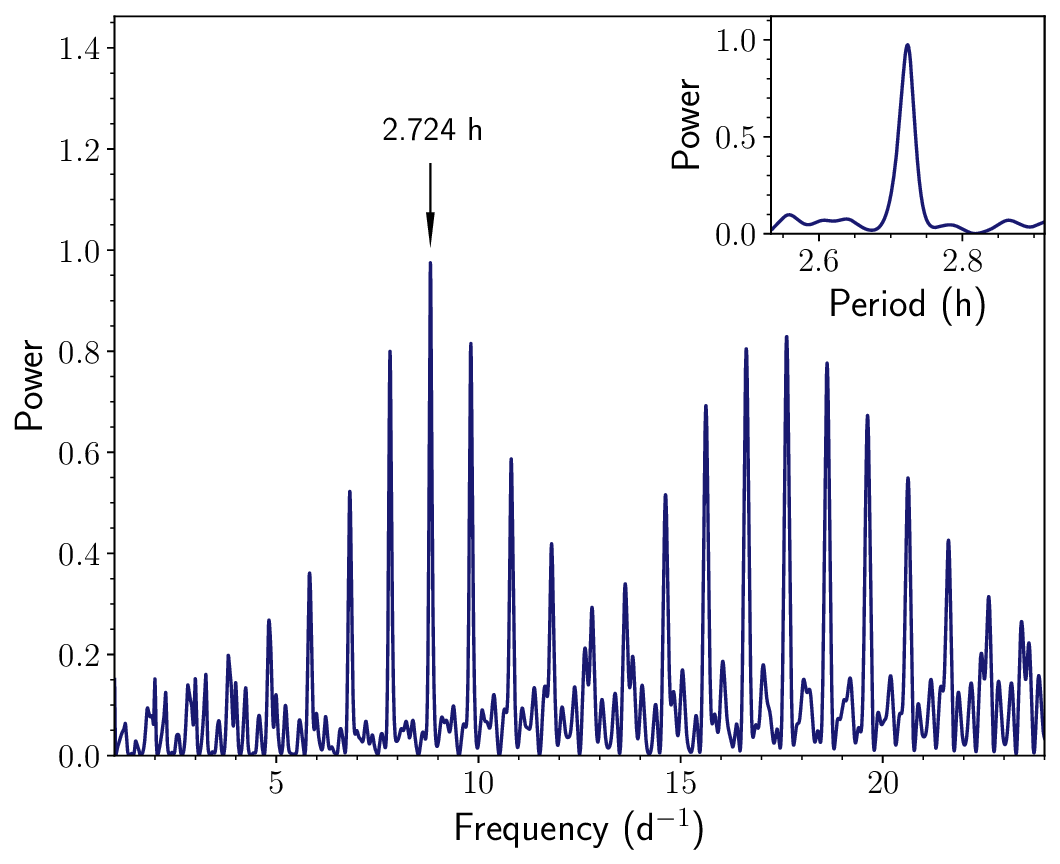}
\end{minipage}
\caption{Lomb-Scargle periodogram. The highest peak corresponding to the best period is enlarged in the inset.}  
\label{fig:LS}
\end{figure}

To search for its periodicities, we performed the Lomb–Scargle periodogram analysis \citep{lomb1976,scargle1982} of the barycenter corrected data in the $R_c$ band where the most numerous  set of data points was obtained, combining the data from six nights. 
The resulting power spectrum is presented in Fig.~\ref{fig:LS}.
The highest peak corresponds to the frequency 8.81 d$^{-1}$, i.e. to the period\footnote{The period uncertainty is derived as the half width at half maximum of the highest peak in the periodogram.
We also used archival ZTF and \ps\ data to better constrain the period. See Appendix~\ref{sec:appendix} for details.} $P_{\rm ph}=2.724(13)$ h.
The second harmonic of the periodicity is also clearly seen in the power spectrum at the frequency 17.62 d$^{-1}$.

\begin{figure}
\begin{minipage}[h]{1.\linewidth}
\center{
\includegraphics[width=1.0\linewidth, clip=]{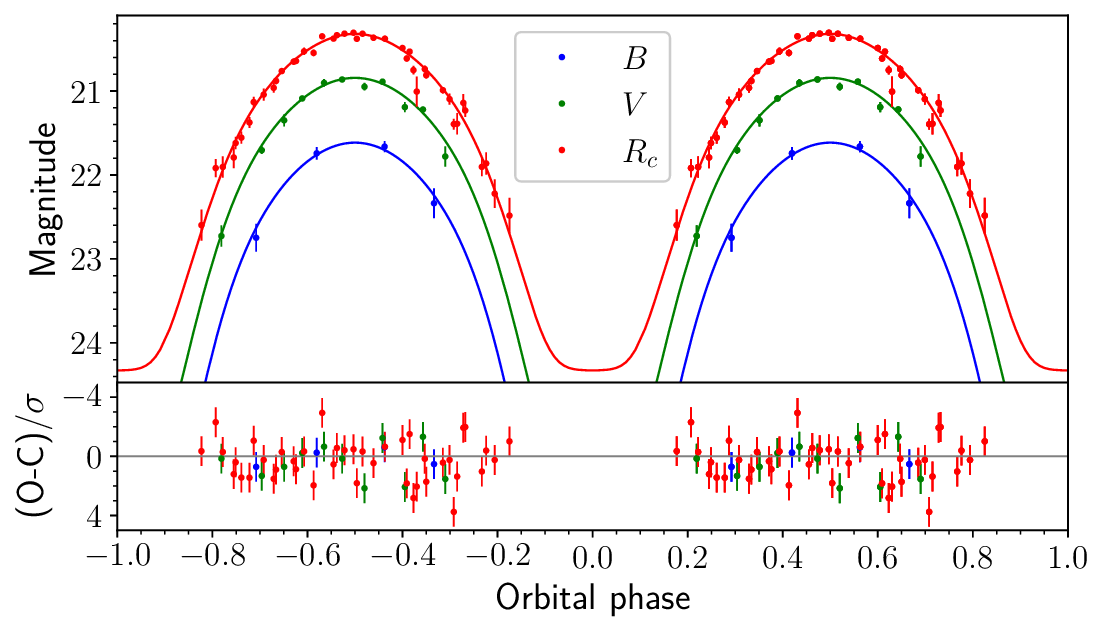}
}
\end{minipage}
\caption{\textit{Top}: Light curves of the \psr\ counterpart candidate obtained with the 2.1-m OAN-SPM telescope and folded with the period 2.724 h. 
The best-fitting model is shown with solid lines.
\textit{Bottom}: Fit residuals derived as the difference between the observed (O) and the calculated (C) magnitudes for each data point in terms of the magnitude
error $\sigma$. }  
\label{fig:curves}
\end{figure}

The $B$, $V$, and $R_c$-band light curves of \psr\ folded with the period are shown in Fig.~\ref{fig:curves}. 
They have a single broad peak per period with a total brightness variation of $\gtrsim 2.5$ mag. Near the minimum brightness phase the source drops below the detection limits of about 23 mag. 
The shapes of the light curves appear to be symmetric and the amplitudes of their brightness  modulation are similar to those observed in other spider MSPs.

%%%%%%%%%%%%%%%%%%%%%%% Fig. Periodogram %%%%%%%%%%%%%%%%%%%%%%%%%%%%%%%
%%%%%%%%%%%%%%%%%%%%%%% Fig. Periodogram %%%%%%%%%%%%%%%%%%%%%%%%%%%%%%%

%%%%%%%%%%%%%%%%%%%%%%%%%%%%%%%%%%%%%%%%%%%%%%%%%%%%%%%%%%%%%%%
%%%%%%%%%%%%%%%%%%%%%%%%%%%%%%%%%%%%%%%%%%%%%%%%%%%%%%%%%%%%%%%

%%%%%%%%%%%%%%%%%%%%%%% Fig. Light curves %%%%%%%%%%%%%%%%%%%%%%%%%%%%%%
%%%%%%%%%%%%%%%%%%%%%%% Fig. Light curves %%%%%%%%%%%%%%%%%%%%%%%%%%%%%%

%%%%%%%%%%%%%%%%%%%%%%% Tab. Fit results %%%%%%%%%%%%%%%%%%%%%%%%%%%%%%%
\begin{table}
\renewcommand{\arraystretch}{1.2}
\caption{The light-curve fitting results for \psr.}
\label{tab:fit} 
\begin{center}
\begin{threeparttable}
\begin{tabular}{lc}
\hline
\multicolumn{2}{c}{Fixed parameters}                       \\ 
\hline
Orbital period, h                                         & 2.724 \\ 
Reddening $E(B-V)$, mag                                   & 0.23 \\

\hline
\multicolumn{2}{c}{Fitted parameters}                       \\
\hline
NS mass $M_{\rm NS}$, \msun                               & $1.96_{-0.60}^{+1.00}$ \\
Mass ratio $q$ =  $M_{\rm c}/M_{\rm NS}$                  & 0.052(5)\\ 
Distance $D$, kpc                                         & 2.13(5)\\

``Night-side'' temperature $T_{\rm n}$, K                 & 3060$_{-60}^{+120}$ \\
Inclination $i$, deg                                      & 83$_{-11}^{+7}$\\
Roche lobe filling factor $f_x$                           & 0.65(5)\\
Irradiation factor $K_{\rm irr}$, \kirr                   & 2.20(14)$\times10^{20}$\\
\hline
$\chi^2$/d.o.f.                                           & 101/50 \\
\hline
\multicolumn{2}{c}{Derived parameters}                     \\
\hline
Companion mass $M_{\rm c}$, \msun                         & 0.102 \\
Companion radius $R_{\rm c,x}$, \rsun                     & 0.21 \\
Companion radius $R_{\rm c,y}$, \rsun                     & 0.14 \\
Lowest ``day-side'' temperature $T_{\rm d}^{\rm min}$, K  & 4010 \\
Highest ``day-side'' temperature $T_{\rm d}^{\rm max}$, K & 7260 \\
\hline
\end{tabular}
\begin{tablenotes}
\tablefoot{\item The Roche lobe filling factor is defined as a ratio of distances from the centre of mass of the secondary to the star surface and to the Lagrange point $L_1$.
\item $R_{\rm c}^x$ and $R_{\rm c}^y$ are the radii of the ellipsoidal companion. The latter is along the line passing through the centres of the binary sources. 
\item Numbers in parentheses denote the 1$\sigma$ uncertainties related to the last significant digits. 
%%%%%%%%%%%%%%%%%%%%%%%%%%%%%%
}
\end{tablenotes}
\end{threeparttable}
\end{center}
\end{table}
%%%%%%%%%%%%%%%%%%%%%%% Tab. Fit results %%%%%%%%%%%%%%%%%%%%%%%%%%%%%%%
   
We fitted the folded light curves with the symmetric direct heating model to derive the parameters of the  presumed spider system.
In this model, the primary is a NS which irradiates and heats a low-mass secondary.
Each surface element of the secondary has a black-body spectrum with an effective temperature varying from element to element. 
For details see \citet{zharikov2013,zharikov2019,kirichenko2024}.

The fitted parameters were the distance to the system $D$, the pulsar mass $M_{\rm p}$, the  mass ratio  of the binary components $q$, the system orbit inclination $i$, the effective irradiation factor $K_{\rm irr}$ defining the companion heating, the companion Roche lobe filling factor $f$, and the ``night-side'' temperature $T_{\rm n}$ of the companion in respect to the pulsar. 
The orbital period was fixed at the measured value $P_{\rm ph}$. 
With $b=22\fdg586$, \psr\ is a high Galactic latitude object. According to the 3D dust map by \citet{dustmap2019}, the interstellar reddening $E(B-V)$ in this direction quickly increases with the distance and reaches  its maximum value of $\approx0.23$ mag at  0.2 kpc. 
The colour and magnitude values of our source suggest that it  is likely a more distant system. 
We thus fixed  $E(B-V)$ at this maximum value.  
The gradient descent method was utilised to find the minimum of the $\chi^2$ function.
The fit results with the 1$\sigma$-uncertainties of the parameters are  presented in Table \ref{tab:fit}, and the best-fitting model is shown with the solid lines in Fig.~\ref{fig:curves}.  
As seen from the figure, the multi-colour light curves are well described by the model. 
The inferred uncertainties are statistical and should be considered preliminary due to the lack of data points.  
For instance, in the absence of data points near the minimum brightness phase, the ``night-side'' temperature value is a prediction  based on extrapolation of the best-fit light curves at phases  around the brightness peak. 
This also affects the resulting inclination angle.   
In addition, there is covariance between the distance and the Roche lobe filling factor. 
An independent distance constraint would be valuable as it can be used as a prior to get a convincing filling factor.

%%%%%%%%%%%%%%%%%%%%%%%%%%%%%%%%%%%%%%%%%%%%%%%%%%%%%%%%%%%%%%%
%%%%%%%%%%%%%%%%%%%%%%%%%%%%%%%%%%%%%%%%%%%%%%%%%%%%%%%%%%%%%%%
%%%%%%%%%%%%%%%%%%%%%%%%%%%%%%%%%%%%%%%%%%%%%%%%%%%%%%%%%%%%%%%
%%%%%%%%%%%%%%%%%%%%%%%%%%%%%%%%%%%%%%%%%%%%%%%%%%%%%%%%%%%%%%%

\section{Discussion}
\label{sec:discussion}

We performed the first multi-band time-series  photometry of the likely optical counterpart to the $\gamma$-ray source \psr. 
We found that its brightness is highly modulated with the period 2.724~h  and an amplitude of $\gtrsim$2.5 mag.  
Its light curves demonstrate a single broad peak per period.
Our findings are consistent with what is observed for spider systems where the heating of the companion face side by the energetic pulsar wind dominates over the tidal distortion of the companion \citep[e.g.][]{draghis2019,kandel2020,matasanchez2023}. 
We thus propose \psr\ as a very promising candidate to such systems. 

The high 2--4 mag amplitude of the \psr\ variability is more typical for BWs than for RBs, whose typical modulation is $\lesssim$1 mag\footnote{Except RB PSR J2339$-$0533, which has the modulation amplitude $\Delta g \approx 6$ mag \citep{kandel2020}.}. 
The presumed orbital period is also more appropriate for BWs: about two dozens of the confirmed members of this family have orbital periods of $<$3 h, while only one RB, PSR J1748$-$2446A\footnote{PSR J1748$-$2446A has $P_b=1.8$~h \citep{lyne1990}. However, it is located in a globular cluster and thus has formation history different to spider pulsars in the Galactic field.}, with such a short period is known \citep{strader2019,swihart2022,psrs-in-gcs}. 
The estimated ``night-side'' temperature of the companion, $\sim$3000 K, is also typical for BWs, which usually have $T_n$ of 1000--3000 K, while RBs are hotter, with $T_n$ of 4000--6000~K \citep{turchetta2023}. 
The difference between the ``night-side'' and the maximum ``day-side'' temperatures is significant, $\Delta T \sim 4000$ K which is again an attribute of BW systems \citep[e.g][]{matasanchez2023}. 
RBs typically have $\Delta T$ of several hundreds Kelvins \citep{koljonen&linares}. 
The Roche lobe filling factor of the \psr\ companion, $f = 0.65$,  is rather low for BWs and RBs, which usually have $f\gtrsim0.8$ \citep[see e.g.][]{strader2019,matasanchez2023}, however, it is not unique.
There are  systems with similar values, e.g. RB J1431$-$4715 ($f=0.73(4)$; \citealt{demartino2024}), BWs J0023+0923 ($f=0.5(1)$) and J2256$-$1024 ($f=0.4(2)$; \citealt{matasanchez2023}). 

The obtained irradiation factor is similar to the values derived for other spider systems \citep[e.g.][]{zharikov2019, kirichenko2024, zyuzin2024}. 

Spider pulsars show a bimodal distribution of the companion  masses with a gap in the range 0.07--0.1~\msun\ between the BW and RB systems \citep{swihart2022}. 
From the light curve modelling, the \psr\ companion has $M_{\rm c}=0.102^{+0.053}_{-0.033}$~\msun, which is consistent with both subclasses. 
The derived mass of the putative pulsar is very uncertain.

According to the \fermi\ Large Area Telescope 14-Year Point Source Catalog \citep[4FGL-DR4;][]{4fgl-dr4}, the \psr\ flux in the 0.1--100 GeV range is $F_\gamma=8.4(7)\times10^{-12}$ \flux.
The \textit{eROSITA} count rate of the \psr\ counterpart candidate is 33.08 cts~ks$^{-1}$ in the 0.2--2.3 keV range \citep{mayer2024}.
Using WebPIMMS\footnote{\url{https://heasarc.gsfc.nasa.gov/cgi-bin/Tools/w3pimms/w3pimms.pl}}, we estimated the unabsorbed flux in the 0.5--10 keV band, $F_X\approx6.2\times10^{-14}$ \flux\ or $\approx 1.5\times10^{-13}$ \flux.
Here we assumed the absorbed power-law spectral model with the photon indices $\Gamma=2.5$ or 1.4, which are the average values for the BWs and RBs, respectively \citep{swihart2022}.
The absorbing column density $N_{\rm H}=2\times10^{21}$ cm$^{-2}$ was derived from the reddening value using the empirical relation of \citet{foight2016}.
For the distance 2.13 kpc obtained from the light curve fitting, the luminosities are $L_X\approx3.4\times10^{31}$ \ergs\ (for $\Gamma=2.5$) or $8\times10^{31}$ \ergs\ (for $\Gamma=1.4$) and $L_\gamma \approx 4.5\times 10^{33}$ \ergs. 
The X-ray-to-$\gamma$-ray flux ratio is $\sim$0.01. 
All these values are typical for spider pulsars \citep{strader2019,swihart2022}.

\section{Conclusions}
\label{sec:conclusion}

Based on our optical and the available $\gamma$-ray and X-ray data, we propose \psr\ as a very promising candidate to a  spider pulsar at a distance of $\approx$2.1~kpc.   
The uncertainties of the derived mass of the putative MSP  companion do not allow us to  determine whether 
it belongs to the RB or BW spider subclasses. 
However, if it is a member of the RB family, it could have the shortest orbital period among the known RBs in the Galactic field.
The model used for the light curve fitting predicts brightness variations with the orbital phase at the levels 
$\Delta B\approx6$ mag, $\Delta V \approx 4.8$ mag and $\Delta R_c \approx 4$ mag. 
Confirmation of this prediction by deeper optical observations near the minimum brightness phase can make \psr\  the second RB with such a strong brightness variability, along with PSR J2339$-$0533 mentioned above.
In the BW case, \psr\ could be among BWs with the most massive companions. 
On the other hand, it might also represent a bridge between these two subclasses.

Targeted searches for periodic millisecond pulsations related to the pulsar spin in the radio and $\gamma$-ray domains are necessary to confirm the spider nature of \psr. 
Detection of the binary radio pulsar J1544$-$2555 using MeerKAT was reported by the TRAPUM (TRansients And PUlsars with MeerKAT) collaboration on their web-site\footnote{\url{https://www.trapum.org/discoveries/}}. 
It has a spin period of 2.39 ms and dispersion measure of 25.8 pc~cm$^{-3}$. 
No other information, e.g. coordinate uncertainties or orbital period, is presented there, however,  most likely it is the same source as the one considered in this work. 
Measurement of the binary period in the radio would  confidently prove  that. 
Optical spectroscopy and multi-band photometry  with a large-aperture telescope would be useful to  measure its radial velocity curve, to detect the source and obtain its colours near the minimum brightness phase and  thus to constrain the parameters of the binary system and the ``night-side'' temperature of the presumed spider companion with a higher precision. 
Deeper studies in X-rays are also encouraged.

\begin{acknowledgements}
We thank the anonymous  referee for useful comments which helped us to improve the paper. 
Based upon observations carried out at the Observatorio Astron\'omico Nacional San Pedro M\'artir (OAN-SPM), Baja California, M\'exico. We thank the daytime and night support staff at the OAN-SPM for facilitating and helping obtain our observations.
This work has made use of data from the European Space Agency (ESA) mission {\it Gaia} (\url{https://www.cosmos.esa.int/gaia}), processed by the {\it Gaia} Data Processing and Analysis Consortium (DPAC, \url{https://www.cosmos.esa.int/web/gaia/dpac/consortium}). Funding for the DPAC has been provided by national institutions, in particular the institutions participating in the {\it Gaia} Multilateral Agreement. 
The Pan-STARRS1 Surveys (PS1) and the PS1 public science archive have been made possible through contributions by the Institute for Astronomy, the University of Hawaii, the Pan-STARRS Project Office, the Max-Planck Society and its participating institutes, the Max Planck Institute for Astronomy, Heidelberg and the Max Planck Institute for Extraterrestrial Physics, Garching, The Johns Hopkins University, Durham University, the University of Edinburgh, the Queen's University Belfast, the Harvard-Smithsonian Center for Astrophysics, the Las Cumbres Observatory Global Telescope Network Incorporated, the National Central University of Taiwan, the Space Telescope Science Institute, the National Aeronautics and Space Administration under Grant No. NNX08AR22G issued through the Planetary Science Division of the NASA Science Mission Directorate, the National Science Foundation Grant No. AST–1238877, the University of Maryland, Eotvos Lorand University (ELTE), the Los Alamos National Laboratory, and the Gordon and Betty Moore Foundation.
Based on observations obtained with the Samuel Oschin Telescope 48-inch and the 60-inch Telescope at the Palomar Observatory as part of the Zwicky Transient Facility project. ZTF is supported by the National Science Foundation under Grants No. AST-1440341 and AST-2034437 and a collaboration including current partners Caltech, IPAC, the Oskar Klein Center at Stockholm University, the University of Maryland, University of California, Berkeley, the University of Wisconsin at Milwaukee, University of Warwick, Ruhr University, Cornell University, Northwestern University and Drexel University. Operations are conducted by COO, IPAC, and UW.
The work of AVK, DAZ and YAS (data reduction, periodicity search) was supported by the baseline project FFUG-2024-0002 of the Ioffe Institute.
DAZ thanks Pirinem School of Theoretical Physics for hospitality. 
SVZ acknowledges DGAPA-PAPIIT grant IN119323. 
AYK acknowledges the DGAPA-PAPIIT grant IA105024.

\end{acknowledgements}


\bibliographystyle{aa}
\bibliography{ref}

\begin{appendix}
\section{Periodicity search adding archival data}
\label{sec:appendix}

\begin{figure}
\begin{minipage}[h]{1.\linewidth}
\center{
\includegraphics[width=1.0\linewidth, clip=]{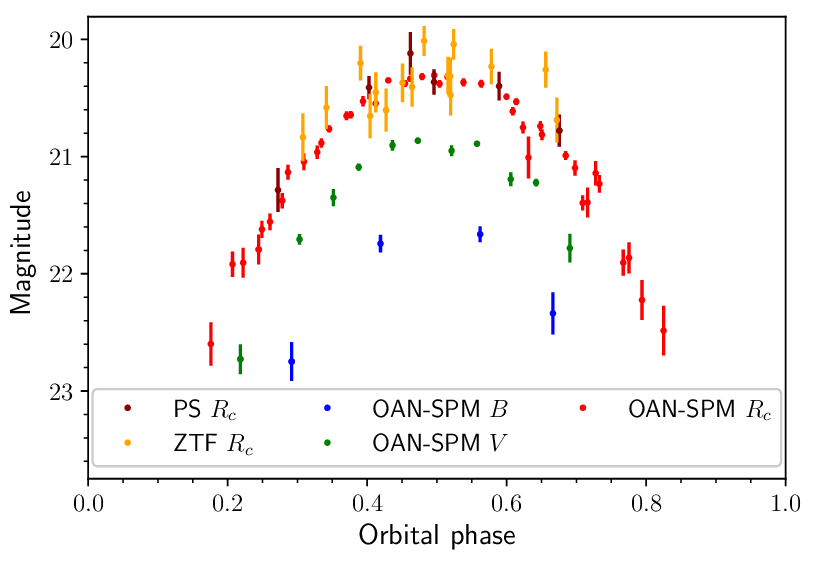}
}
\end{minipage}
\caption{Light curves of the \psr\ counterpart candidate folded with a period of 2.723884 h. 
The data from different instruments/filters are marked by various colours as indicated in the legend (PS $\equiv$ \ps).}  
\label{fig:lc-app}
\end{figure}

We performed periodicity search  adding available $r$-band data from the Pan-STARRS DR~2 (6 data points) and ZTF DR~22 (16 data points) catalogues to the OAN-SPM $R_c$-band data. The Legacy Surveys DR 10 does not have measurements in this band. 
The archival data are sparse, but they allowed us to cover a time range of about 11 yr.
We transformed the archival magnitudes into the $R_c$-band magnitudes using equations from \citet{tonry2012}. Searching for  periodicity using the combined data set resulted in the largest peak in the power spectrum near the same frequency as without the archival data. To obtain the best period value and its uncertainty, we simulated 2000 light curves with magnitudes scattered around the central values of the measured points, assuming a normal distribution with the standard deviation determined by the magnitude uncertainty.
Times of measurements were uniformly distributed within the exposure times of individual frames.
To each simulated light curve we applied the Lomb-Scargle periodogram method and obtained the period distribution. 
The mean value and standard deviation of the latter were taken as the best period and its uncertainty, $P_{\rm ph} = 2.723884(1)$~h. This $P_{\rm ph}$ is consistent with  that derived in Sec.~\ref{sec:lc-mod} using only  the OAN-SPM data, while its  uncertainty is much smaller due to the much longer time interval covered by the archival data.
The light curves folded with this period are shown in Fig.~\ref{fig:lc-app}.
One can see that the archival data  are in agreement with the OAN-SPM data although the Pan-STARRS and ZTF surveys were able to detect \psr\ only near its maximum brightness phase and with significantly higher magnitude uncertainties.

%%%%%%%%%%%%%%%%%%%%%%% Fig. Light curves %%%%%%%%%%%%%%%%%%%%%%%%%%%%%%
%%%%%%%%%%%%%%%%%%%%%%% Fig. Light curves %%%%%%%%%%%%%%%%%%%%%%%%%%%%%%

\end{appendix}

\end{document}